\documentclass[a4paper,fleqn]{mnras}

\usepackage[T1]{fontenc}
\usepackage{ae,aecompl}

\usepackage{graphicx}	

\title[Increasing Geminid meteor shower activity]{Increasing Geminid meteor shower activity}

\author[G. O. Ryabova and J. Rendtel]{
G.~O.~Ryabova$^{1}$\thanks{E-mail: goryabova@gmail.com} and
J.~Rendtel$^{2}$
\\
$^{1}$Research Institute of Applied Mathematics and Mechanics of Tomsk State University, 
Lenin pr. 36, 634050 Tomsk, Russian Federation\\
$^{2}$Leibniz-Institut fur Astrophysik Potsdam (AIP), An der Sternwarte 16, 14482 Potsdam, Germany\\
}

\date{Accepted XXX. Received YYY; in original form ZZZ}

\pubyear{2017}

\begin{document}
\label{firstpage}
\pagerange{\pageref{firstpage}--\pageref{lastpage}}
\maketitle

\begin{abstract}
Mathematical modelling has shown that activity of the Geminid meteor shower should rise with time, and that was confirmed by analysis of visual observations 1985--2016. We do not expect any outburst activity of the Geminid shower in 2017, even though the asteroid (3200) Phaethon has a close approach to Earth in December of 2017. A small probability to observe dust ejected at perihelia 2009--2016 still exists.
\end{abstract}


\section{Introduction}

The Geminid meteor shower is an annual major shower with the maximum activity near December 14. In 2017 asteroid (3200) Phaethon (the parent body of the stream) has a close encounter with the Earth on December 16, and the minimum distance was calculated to be 0.0689~au (Galushina, Ryabova \& Skripnichenko 2015).

When a comet approaches the Sun it is expected that its meteor shower (if it exists) will display enhanced activity. The young stream, ejected from the comet on the previous revolution, had insufficient time to spread around the orbit, so the swarm located in the vicinity of the nucleus can produce outburst meteor activity on the Earth. However, this is not necessarily the case, and with high probablility is not applicable to the Phaethon--Geminid complex.

A mathematical model, which has been developed over the last twenty years (see (Ryabova~2007, 2016) and references therein), gives the following scenario of the Geminid stream formation. About two thousand years ago comet Phaethon was captured on an orbit with perihelion distance 0.10--0.12~au. The catastrophic release of particles and volatiles caused a dramatic transformation of the orbit. The Geminids were generated during a short time and had no replenishment after that. This scenario does not imply an increase of the Geminid shower activity during Phaethon's encounter with the Earth. Phaethon approaching the Earth does not mean that the Geminid stream core approaches the Earth. 

Analysis of 60 years of visual observations (1944--2003) has shown that the shower activity is rather stable. Meanwhile, the modelling fulfilled in anticipation of the Phaethon encounter has shown that the shower activity should grow. The aim of our research was to revisit the previous analysis of the visual observations (Rendtel~2004) adding 13 years of observations made since then, to compare the results with the modelling, and to explain why we expect the increase in activity.

\section{Model}

\subsection{Activity}

Ryabova (2016) presented the Geminid meteoroid stream models, consisting of 30~000 meteoroids with fixed masses (0.02, 0.003 and 0.0003~g). The stream was generated around starting epoch JD~1720165.2248 (perihelion passage) using the cometary scenario of ejection. For details of the model, method and references see (Ryabova 2016). We used one of these models with meteoroids of the `visual' mass of 0.02~g and extended it until 2025 January~01. We tracked the encounters with the Earth assuming that model meteoroids approaching the Earth to a distance $\leq$~0.02~au are recorded at the Earth.

\begin{figure}
\begin{center}
\includegraphics[width=70mm]{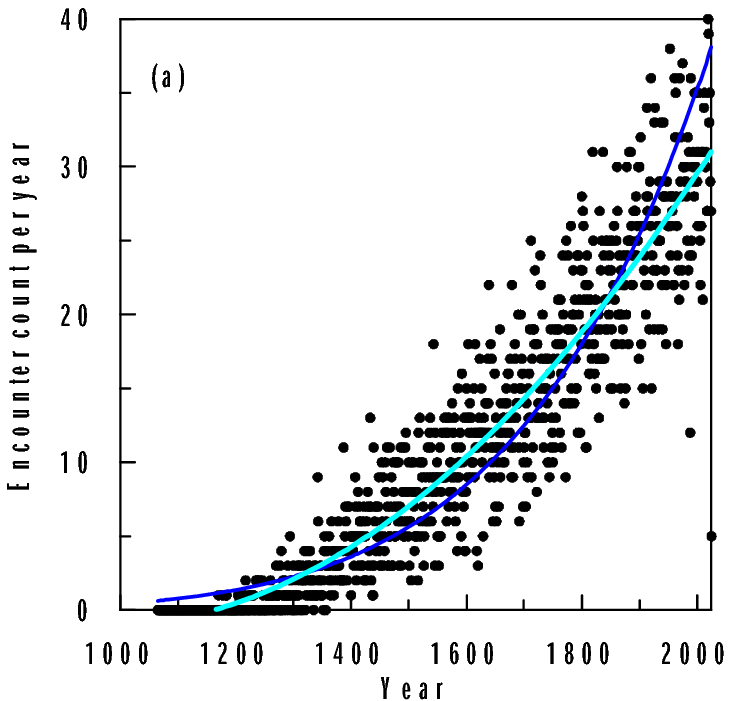}

\includegraphics[width=70mm]{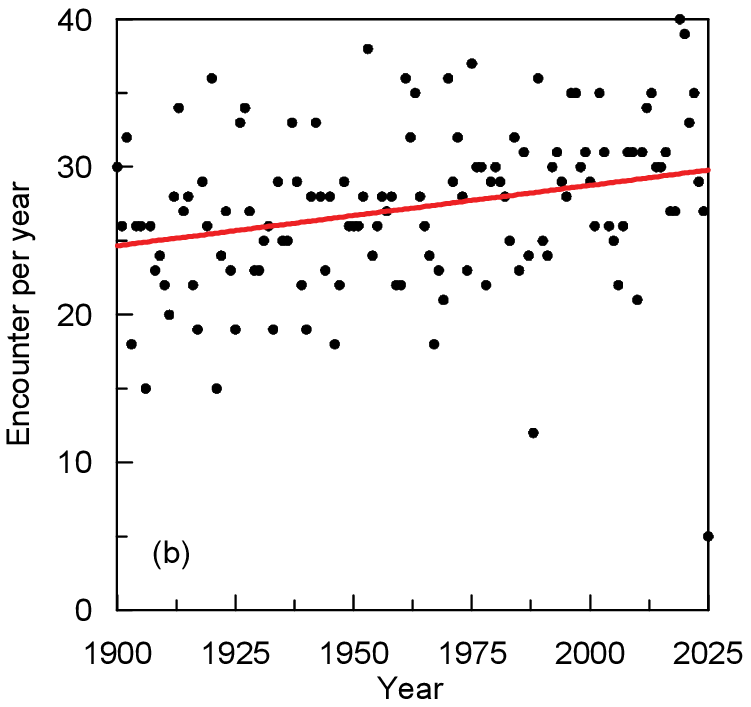}
\caption{
Model meteoroid encounters (up to 0.02~au) with the Earth as a function
of time. (a) For the 
entire period of the stream's existence. The cyan line shows the polynomial fit (\ref{eq:02}), 
and the blue line shows the power fit (\ref{eq:03}). (b) For XX and XXI centuries. The red 
line shows the linear fit (\ref{eq:01}).
}
\label{fig:01}
\end{center}
\end{figure}

Fig.~\ref{fig:01} demonstrates that the encounter rate increases. In the last 125 years it can be approximated by the linear equation
\begin{equation}
N_{\rm enc} (yr) = 0.041 \times yr - 53.2,
\label{eq:01}
\end{equation}
\noindent
where $N_{\rm enc}$ is the encounter rate, and $yr$ is the year. However the complete period is fitted best by the polynomial or the power law fit:
\begin{equation}
N_{\rm enc} (yr) = 25.93 - 0.056 \times yr + 2.88 \times 10^{-5}yr^2,
\label{eq:02}
\end{equation}
\begin{equation}
N_{\rm enc} (yr) = 3.18\times 10^{-20}yr^{6.375}.
\label{eq:03}
\end{equation}

\subsection{Why activity increases}

\begin{figure}
\begin{center}
\includegraphics[width=70mm]{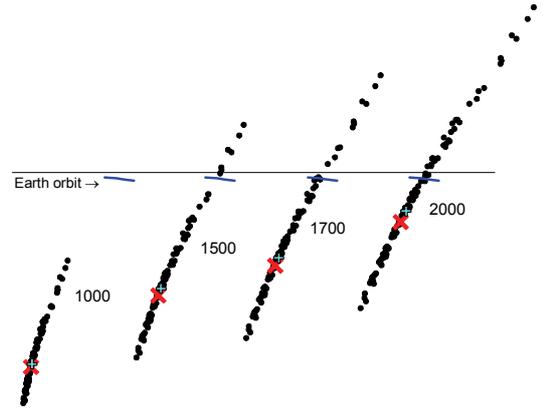}
\caption{
The model Geminid stream (100 particles, mass = 0.02~g) cross-sections in the
ecliptic plane at the descending node in the years 1000, 1500, 1700 and 2000. Nodes of the meteoroids are designated by black dots. The small cyan cross (+) is the node of Phaethon's orbit, the large red cross is the Geminid's mean orbit node, the blue line is the Earth orbit. The horizontal line shows that the cross-sections are aligned. We used the standard heliocentric ecliptic system as
reference system.
}
\label{fig:02}
\end{center}
\end{figure}

The model Geminid stream undergoes strong anisotropic dispersion (Ryabova 2007,
section 3.1 and fig.7). Its intersection with the ecliptic plane progressively
stretches (Fig.~\ref{fig:02}) and at the same time it moves away from the Sun. Phaethon's node and the mean orbit of the stream (i.e. the densest part of the
stream) gradually approach the Earth's orbit. So the Geminid shower activity slowly increases. 
We should mention that the Poynting--Robertson drag makes meteoroids (but not the
asteroid) spiral slowly towards the Sun, hence the node of the mean stream orbit (= the densest part or the stream core) also shifts towards the Sun. That is why the cyan cross separates from the red cross in Fig.~\ref{fig:02} with time. However,
the drift in the evolution from gravitational effects is much stronger.

Phaethon's node should intersect the Earth's orbit about 2200, and the model Geminid stream core some time later. After that the Geminid shower activity should decrease. In 2200 the activity could reach 140~per~cent of its current rate according to fit (\ref{eq:02}). 

Studying the evolution of the minimal distance between the nominal orbit of Phaethon and the Earth's orbit Jakub\'\i k \& Neslu\v{s}an (2015) have already implied that `the activity of this shower is currently increasing' (i.e. Geminid shower). In essence this is correct, but the process is more complicated, as we see.  

\section{Optical observations of Geminids 1985--2016}

The present analysis differs from the previous one (Rendtel 2004) in three essential features. Firstly, 
here we use only homogeneous observations. The global data used for this study are stored in the Visual Meteor Data Base (VMDB) of the International Meteor Organization (IMO). It is available from the website {\tt www.imo.net}. 
The analysis presented in Rendtel (2004) ends with the 2003 return. Right now we can add another 13 years at the end of the series. 

The second difference is that here we concentrate on the temporal evolution of the Geminid peak activity.
So we calculated Zenithal Hourly Rate (ZHR) only for the peak period between $261^\circ-262^\circ$ in solar longitude.

Thirdly, we used a constant population index $r=2.4$. Since we are interested in the long-term evolution, we omitted a separate $r$ determination per return. This last point needs some explanation.

The ZHR is the number of shower meteors $N$ seen per hour by one observer  
under standard conditions, i.e. radiant in the zenith (elevation 
$h_R=90^\circ$), limiting stellar magnitude ${\rm LM} = +6.5$~mag and unlimited field 
of view. The latter translates into an effective field of view of $52^\circ$ 
radius (Koschack \& Rendtel 1990). For the correction, the population index $r$ 
describing the increase of the number of meteors in subsequent magnitude 
classes is essential:
\begin{equation}
{\rm ZHR}=\frac {c_F N r^{(6.5-{\rm LM})}} {T \sin h_R},
\label{eq:04}
\end{equation}
\noindent
with $c_F$ being the correction for the geometrical field obstruction. Hence
the knowledge of the population index $r$ needs to be derived for each
return of the shower. It has been found that $r$ varies during the Geminid 
activity period. For most of the activity period of the Geminids we find $r=2.6$, 
but towards the peak it decreases to $2.5 - 2.1$. Detailed analyses have 
been published for the well documented returns in 1993 (Arlt \& Rendtel 1994), 
1996 (Rendtel \& Arlt 1997) and 2004 (Arlt \& Rendtel 2006).
So for our search of the peak ZHR we applied $r=2.4$ which is
typical for the main peak section of the activity profile.

\begin{figure}
\begin{center}
\includegraphics[width=80mm]{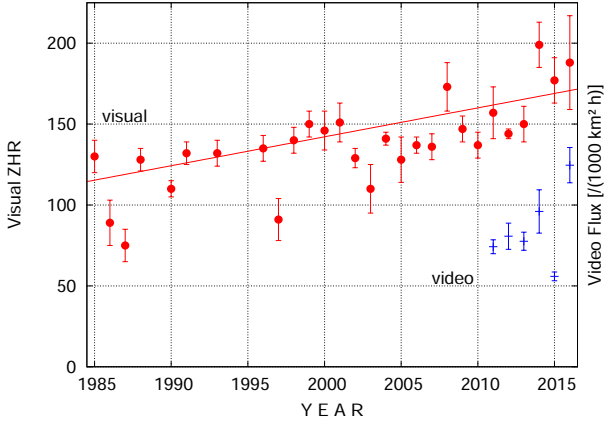}
\caption{
Activity level of the Geminids in the period 1985--2016. We show the peak 
value of the visual ZHR for each year in the period $261^\circ - 262^\circ$ in solar longitude
as well as the maximum flux obtained from the IMO Video Meteor Network data since 2011. The line shows a fit of
the visual ZHR derived for the period 1985--2016.
}
\label{ZHR}
\end{center}
\end{figure}

The resulting ZHR for all returns is shown in Fig.~\ref{ZHR}. 
The increase of the peak ZHR with time is obvious in the interval under study 
and can be described by a fit like
\begin{equation}
{\rm ZHR}(yr) = 1.79 \times yr-3431.41.
\label{eq:05}
\end{equation}

Observations under poor conditions, such as bright moonlight, may suffer from a small sample and from an under-correction. The latter was found in several analyses of different meteor showers and is probably due to perception changes. In fact, the Geminid returns of 1997, 2003 and 2005 (below the fitted line) occurred
with bright moonlight. Other low values (1986, 1987, 2015) are not affected in the same way (no moonlight).
Another example is the 2008 return which coincided with the full Moon, but deviates in the opposite direction.
This suggests rather fluctuations between individual stream encounters and needs further investigation.
The lower ZHR in 2015 found in both video and visual data independently also supports that we are able to detect real stream features.

Another supporting and independent data sample is the collection of video 
flux data derived from the data stored by the IMO Video Meteor Network 
(Molau \& Barentsen 2014). The flux can be calculated via the access to 
{\tt meteorflux.io} for each year since 2011. We added the respective flux 
values given as number of meteors intersecting the normal area in 1000~km$^2$ per hour. 
The peak values of the visual and video data show the same pattern, supporting that the 
variations are not caused by observing conditions or selection effects
but are real.

\section{Discussion}
\subsection{The previous and the current analyses of observations} 

Despite the fact that the Geminids are currently the most active meteor 
shower, it gained rather little attention in the past. Like with
many other meteor showers, our knowledge of the activity level more than
about 30--40 years back is scarce. So our previous analysis for the Geminids (Rendtel 2004) was based on two kinds of observations. Several series were taken from the literature and re-processed to find ZHR (1944--1984). Other series (1985--2003) were taken from IMO VMDB, and they were collected and processed according to the standardized procedure. The Geminid shower activity was found to be rather stable. A slight increase was noticed and mentioned, but within the error margins. For the new analysis we used only homogeneous series of visual observations from VMDB (1985--2016). 

The previous analysis also included data for all the shower activity period. That led to the large scatter, and the trend in ZHR had substantial noise. Confining ourselves to the peak period and constant population index $r$, we eliminated some part of this noise, so the activity increase became apparent. Only during the last 17 years (i.e.\ 2000--2016) activity has increased by 20~per~cent!

\subsection{Modelling and observations}

Model shower activity increase is obvious in Fig.~\ref{fig:01}a and not so obvious in Fig.~\ref{fig:01}b. During 2000--2017 the activity growth comes to only 2.3~per~cent, if we use the linear fit (\ref{eq:01}), 3.2~per~cent for the power fit (\ref{eq:03}) and 5.2~per~cent for the polynomial fit (\ref{eq:02}). Why is there such a striking difference with observations?

The quality of the fit (\ref{eq:01}) is not good (coefficient of determination here is only 0.07), and the reason could be too small an amount of data points used (namely, 126). However for equations (\ref{eq:02}) and (\ref{eq:03}) the coefficients of determination are 0.89 and 0.83, respectively. So the reason of the difference is not the quality of the fit, but the model itself. The model by Ryabova (2007, 2016) is a {\it qualitative} model. It has two significant (and probably unremovable) discrepancies --- in the location and in the width --- with the real stream. The discrepancies and their reason were discussed in detail in (Ryabova 2007, 2016), and also in earlier papers. Being a qualitative model it shows trends and patterns in the stream behaviour and structure, and explains them. It is not intended for numerical estimations of the stream parameters though.

\subsection{Could the `thermal' Geminids be observed in 2017?}

Phaethon was discovered in 1983 (as 1983TB) and no activity was observed till in 2009 Phaethon brightened by 2~mag just after perihelion (Jewitt \& Li~ 2010). The similar two-days brightening around perihelion was also observed in 2012 (Li \& Jewitt~2013) and 2016 (Hui \& Li~2017). The most plausible reason of this phenomenon is an ejection of dust generated by thermal fracture (Jewitt \& Li~ 2010). Very soon it became clear that these perihelion outbursts are not the main component of the stream (Li \& Jewitt~2013, Ryabova~2015, Hui \& Li~2017). Ryabova (2012) found that 0.03~per~cent of the model dust swarm ejected in 2009 approached the Earth to distances 0.018 -- 0.03~au  in 2017. We made similar modelling (using the techique described in (Ryabova~2012)) for ejections in 2012 and 2016. In all three cases particles approach the Earth at solar longitudes 262\fdg45$\pm$0\fdg005 to distances down to 0.019~au. The radius of the Earth's sphere of influence is about 0.03~au, so considering possible variations in ejection models and additional gravitational scattering close to Earth, there is a small probability that these meteoroids reach Earth to become observable. It is difficult to estimate whether we can distinguish them from the `regular' Geminids.

\subsection{The previous Phaethon close encounters with the Earth}

In the XX century, Phaethon had two close encounters with the Earth: 1931 December 13 (0.03839~au) and 1974 December 16 (0.05474~au). Though we believe that the Geminid activity does not depend on  Phaethon's position, it is worth checking the old observations.
 
The situation turned out to be unpromising for both
times. There are descriptions of single observations in 1933 and 1934 
which do not allow us to derive an activity level (Millman 1934, 1935).
This also holds for reports of some 1974 Geminid visual observations compiled in 
the magazine Sky \& Telescope (anon.\ 1975, p.~194). A calibration is 
not possible because we have information neither about the observing 
conditions nor on activity observed in the adjacent years nor on the number
of sporadic meteors recorded during these observations. The fact that 
a high activity was mentioned may mean that the maximum itself was not observed in other years. There may be
still treasures in some archives. 

The last point applies also to the asteroid (3200) Phaethon. Now, when many astronomical archives are digitized and some useful tools for old asteroid observation searches, like S{\sevensize KY}B{\sevensize O}T (Berthier et al. 2016, IMCCE VO 2017) are accessible for the research community, there is a chance to find pre-discovery observations of Phaethon.

\section{Concluding Remarks}

We analysed visual observations of the Geminid shower in 1985--2016 around the shower maximum using homogeneous series of visual observations. It was found that the shower activity slowly increases. The same was obtained for video observations (2011--2016). These results were supported and explained by mathematical modelling. Activity of the shower increases because the core of the Geminid stream moves towards the Earth.

A small probability exists that particles ejected in 2009 around perihelion of Phaethon's orbit could be observed at solar longitudes 262\fdg45$\pm$0\fdg005 (J2000.0).

We do not expect any outburst Geminid activity in 2017, but only a wide observational campaign could validate or disprove our {\it theoretical} expectations.

\section*{Acknowledgements}
G.R. was supported by Ministry of Education and Science of the Russian Federation project No. 9.9063.2017/BCh. 
This research has made use of NASA's Astrophysics Data System. The authors
thank the anonymous referee, who put forward several very relevant questions. We are grateful to David Asher for reading and discussing the manuscript.

\label{lastpage}
\end{document}